\documentstyle[aps,twocolumn]{revtex}
\begin{document}
\draft
\wideabs{
\title{Critical-current density from magnetization loops of finite
high-$T_c$ superconductors}
\author{Alvaro Sanchez$^1$ and Carles Navau$^{1,2}$}
\address{$^1$ Grup d'Electromagnetisme,
Departament de F\'\i sica, Universitat Aut\`onoma Barcelona \\
08193 Bellaterra (Barcelona), Catalonia, Spain\\
$^2$ Escola Universit\`aria Salesiana de Sarri\`a, Rafael
Batlle 7, 08017 Barcelona, Catalonia, Spain}
\maketitle
\begin{abstract}
We analyze the effects of demagnetizing fields in the magnetization
hysteresis loops of type-II superconductors, by a model that allows
the calculation of current and field profiles in finite cylindrically
symmetric superconductors in the critical state. We show how the
maximum in the magnetization curve gradually shifts from negative
applied field values to the central position with decreasing sample
thickness. From the analysis of the calculated field profiles, we
demonstrate that one can obtain the intrinsic field-dependent
critical-current density of the superconductor with great accuracy by
measuring the magnetic response of superconductors with large aspect
ratio with the applied field parallel to the shortest dimension.
\end{abstract}
}

The critical current density $J_c$ is one of the key parameters of
high-$T_c$ superconductors. In particular, knowing the dependence of
$J_c$ on the internal magnetic field $H_{\rm i}=B/\mu_0$ in the
superconductor is a key concern for the study of the current-carrying
states of the superconductors. The $J_c(\vert H_{\rm i}\vert)$
dependence is obtained from two general types of experiments:
electrical and magnetic. In electrical transport measurements, the
reliability and repeatability of the measured values of the critical
current is reduced because of the difficulties in making contacts in
the sample and in choosing a voltage criteria for defining $J_c$.
Moreover, in general large currents circulate through the sample
during the measurements, so that they produce a magnetic field that
modifies the applied field in a inhomogeneous way. This makes the
estimation of the intrinsic $J_c(\vert H_{\rm i}\vert)$ dependence
difficult. The other general method for obtaining $J_c$ is from
magnetization measurements, typically isothermal magnetization loops
$M(H_{\rm a})$, where $H_{\rm a}$ is the applied field.  $J_c$ is
obtained from the width $\Delta M$ of the $M(H_{\rm a})$ hysteresis
loop using 
\begin{equation}
J_c(H)=\Delta M(H)/d, 
\end{equation}
where $d$ is a length
characteristic of the sample size and geometry; for a cylinder of
radius $R$,
\begin{equation}
J_c(H)=3\Delta M(H)/2R. 
\end{equation}
This relation follows from the
critical-state model \cite{bean}. The magnetic method also involves
some approximations. First, the method was originally derived for the
case of $J_c$ independent of $\vert H_{\rm i}\vert$\cite{bean}. If
instead $J_c$ depends on $\vert H_{\rm i}\vert$, as in most actual
superconductors, what is obtained from the width of the loop is not
the required intrinsic $J_c(\vert H_{\rm i}\vert)$ function, but a
different function, $J_c(\vert H_{\rm a}\vert)$. These functions are
approximately equal only if some extra conditions concerning internal
field homogeneity are met, as discussed in
\cite{fietz,chengold,chaddah,johansen} and also below. The second
important approximation is that the method is only theoretically
justified for infinitely long samples. The validity of the method for
extracting $J_c$ from the width of the loop is therefore questionable
in the realistic case of finite superconductors. 

Calculations of the magnetic response of finite superconductors in
the critical state including demagnetizing effects have been recently
presented for strips\cite{brandtstrip} and cylinders
\cite{brandtdisk,doyle,ieee}, following previous works on very thin
strips \cite{thinstrips} and disks \cite{thindisks}. Very recently,
Shantsev et al \cite{shantsev} have discussed important features of
the effects of demagnetizing fields in the hysteresis loop. In
particular, they demonstrated theoretically and experimentally that
for very thin strips in perpendicular field the peak that appears in
the reverse
magnetization curve is located not at negative fields, but at the
central position.

In this work, we systematically study the effect of demagnetizing
fields in the magnetic response of finite superconducting cylinders
in the presence
of a uniform applied field, and discuss the implications in $J_c$
extraction. We first introduce a model which allows us to compute
current and field profiles and magnetization loops of superconducting
cylinders with the same intrinsic parameters but different aspect
ratios, in order to study in detail the effects of demagnetizing
fields. These will be analyzed in
relation to two particular features: the position of the peak in the
magnetization curves as a function of the sample aspect ratio,
extending the work of Shantsev et al \cite{shantsev}, and the
relation between the shape of the hysteresis loop and the field
dependence of the critical current density for the different aspect
ratios. 

Our model, which simulates the process of penetration of
supercurrents inside a superconductor, is based in the fact that
any modification of the applied field results in a change in the
superconductor current distribution in order to minimize the change
in the magnetic energy. The current distribution in the initial
magnetization curve (after the sample has been zero-field cooled and
a magnetic field is applied) can therefore be obtained from the
magnetic energy minimization. We will obtain the reverse current
distributions (once the field has reached a maximum value and is
decreased) by the conventional procedure of superposing a current
distribution with opposite sign to the 'frozen' field profiles as in
\cite{thinstrips}. In all cases, we assume that $B=\mu_0 H$, which
holds for fields $H_{c1}<< H << H_{c2}$, as is usually
done in critical-state modeling.

Consider a cylindrical type-II superconductor of radius $R$ and
length $L$ located in a uniform applied field, $H_{\rm a}$, directed
along its
axis. We use common cylindrical coordinates ($\rho,\theta,z$), $z$
being the direction of the axis of the superconductor. Owing to the
symmetry of the system (which reduces the problem to a
two-dimensional one), all supercurrents flow with angular direction.
We divide the superconductor in a regular grid of $n \times m$
coaxial rings in which linear currents can flow. The magnetic flux
that threads one of these linear circuits at the position ($\rho,z$)
due to the external applied field is $\Phi^a(\rho,z)=\mu_0 H_{\rm a}
\pi \rho^2$, while the flux that threads the same circuit due to all
currents circulating in the superconductor is
$\Phi^i(\rho,z)=\sum_{\rho',z'}M(\rho,z,\rho',z') I(\rho',z')$, where
$I(\rho',z')$ indicates the current in the circuit ($\rho',z'$), and
$M(\rho,z,\rho',z')$ the mutual inductance between the circuits
($\rho',z'$) and ($\rho,z$). The self-inductances $M(\rho,z,\rho,z)$
are calculated from the mutual inductance between two close linear
circuits \cite{ieee}. The model details are described in Refs.
\cite{ieee,araujo}. In this paper we have used typical values of $n
\times m \simeq 120\times20$ for thin samples and $n \times m \simeq
60\times60$ for samples with larger $L/R$ ratios.

Let us assume we have a given
current distribution corresponding to an applied field $H_a$ (if we
are calculating the first point after the initial state, then the
initial current distribution is zero everywhere). Setting a
current $I$ at a circuit $(\rho,z)$ requires
an energy $E=I\Phi^i(\rho,z)$
while it contributes to reduce the energy (current has opposite sign
to $H_a$) by a quantity $I \Phi^a(\rho,z)$. We find in this way the
circuit that yields the largest decrease of energy and set a current
$I$ there. The new currents are set accomplishing the chosen material
law $J=J_c(\vert H_{\rm i} \vert )$, $\vert H_{\rm i}\vert$ being
the modulus of the
total field ${\bf H}_i$, and $I=J_c(\vert H_{\rm i} \vert ) (RL/nm)$.
When no new currents minimize further the energy, we
calculate the magnetic field inside the superconductor and change the
value (not the distribution) of the already induced currents to
accomplish the material law. 
At this point, we calculate the magnetic moment resulting from all
the circulating currents and all the other relevant magnitudes. After
this, the applied
field can be increased again, and the
process is restarted from the existing current distribution.
The reverse stage (corresponding to decreasing $H_a$ from $H_{\rm
max}$ to $-H_{\rm max}$) is calculated by superposing the frozen
current penetration set at $H_{\rm max}$ to the one induced in the
reverse stage (calculated in a similar manner as described for the
initial curve). This procedure is typical of the critical state model
\cite{thinstrips}.

Our model allows the implementation of an arbitrary $J_c(\vert H_{\rm
i}\vert)$ dependence. The following discussions and conclusions are
valid for any dependence as long as $J_c$ is a decreasing function on
$\vert H_{\rm i}\vert$, which is physically reasonable. For
illustrating our results, we choose an exponential dependence
$J_c=J_{c0}\exp(-\vert H_{\rm i}\vert/H_0)$, where $J_{c0}$ and $H_0$
are positive constants. The exponential dependence has
been successfully applied to high-$T_c$ superconductors
\cite{expmod,physc}. A useful parameter for the analysis is
$p=J_{c0}R/H_0$, which takes on large values for a strong dependence
of $J_c$ on $\vert H_{\rm i}\vert$ and tends to 0 for $J_c$
independent of field \cite{expmod}. Realistic values of $p$ for
high-$T_c$ superconductors range from 1 to 10 \cite{physc}. 

In Fig. 1 we show the calculated $M(H_{\rm a})$ curves for different
values of $L/R$, for the cases $p=0$ (Bean's model), 3, and 10. $M$
and $H_{\rm a}$ are normalized to $H_{\rm p}=H_0\ln(1+p)$, which
corresponds to the penetration field for an infinite cylinder. We
find that the calculated $M(H_{\rm a})$ loops for sufficiently large
samples coincide between our numerical accuracy with the known
analytical results for infinite cylinders\cite{expmod}. Let us now
discuss the differences observed in the loops for each value of $p$,
which are only due to the sample geometry. In all cases, the
demagnetizing field enhances the initial slope of both the initial
and
reverse curves. We have analyzed this effect in detail, observing
good agreement with experimental data measured for niobium cylinders
of different lengths in \cite{araujo}; we obtain the same values for
the initial slope of the $M(H_{\rm a})$ curves as those calculated by
Brandt \cite{brandtdisk} and Chen et al \cite{chendemag} for a wide
range of $L/R$ values, with less than 1\% deviation.  The shape of
the central part of the loop is not changed in the Bean case ($p=0$),
since even in a finite sample the width of the loop is proportional
to the the constant value of $J_c$. However, important variations are
observed for non-constant $J_c$. When $p\neq 0$, a peak appears in
the reverse magnetization curve. In Fig. 2 we show the calculated
dependence of the peak position $H_{\rm peak}$ on $L/R$ for $p$=3 and
10.

The general trend of the peak position is that it tends towards
$H_a=0$ with decreasing sample $L/R$ ratio.
A similar tendency was found in \cite{zphys} for a three
parameter $J_c(H_i)$ function and using an iterative method valid
only for the fully penetrated stage.
Our calculated results give the correct known
limit when the sample is very large (solid line in Fig. 2
indicates the limit for infinite cylinders \cite{expmod}).  Our
results are also compatible with \cite{shantsev2}, where it is said
that the
peak tends towards $H_a=0$ for thin samples without reaching exactly
the central position (this is in contrast with the strip case, for
which the peak position was shown to be zero for very thin strips
\cite{shantsev,mcdonald}).

The peak position and the other observed features of the hysteresis
loop can be understood as follows. We start by discussing the
situation in the simpler case of an infinitely long superconductor
with a given $J_c(\vert H_{\rm i}\vert)$ function. After reaching the
maximum applied field $H_{\rm max}$, when we reverse the magnetic
field sweeping direction in the hysteresis loop, supercurrents are
induced at the surface of the cylinder with a direction opposite to
the shielding currents of the initial magnetization curve. These
reverse supercurrents gradually enter the sample, confining the
original currents to the interior. When $H_{\rm a}$ decreases, the
internal field in the region penetrated by reverse currents also
decreases, these currents become larger, and therefore the
magnetization increases. Passing through $H_{\rm a}=0$, the internal
field becomes negative at the surface regions of the superconductor.
Further decreasing $H_{\rm a}$ makes the $H_{\rm i}$ profile
negative, so that $\vert H_{\rm i}\vert$ again increases. Since $J_c$
decreases with the absolute value of $H_{\rm i}$, the currents become
low and $M$ decreases. Then, a peak appears in the magnetization at
some negative value of $H_{\rm a}$, for which an averaged value of
$\vert H_{\rm i}\vert$ is minimum. This process is depicted in Fig.
3a, where we show the calculated field profiles in the cylinder
midplane corresponding to the case $p=10$ for a long cylinder with
$L/R=10$, which schematically represents the infinite case. 

There is a key difference for this case and the behavior observed for
a realistic finite sample. Whereas in an infinite sample the
internal field at a given point in the superconductor has only
contributions from $H_{\rm a}$ and the field created by the currents
exterior to this point \cite{chengold,expmod}, there is contribution
from all the circulating currents at all points in a finite
superconductor. This self-field contribution produces an effect in
the central region of the loop ($\vert H_{\rm a}\vert$ small) similar
to shifting the applied field upwards to higher values for $H_{\rm
a}>0$ and downwards to lower values for $H_{\rm a}<0$. This explains
the increase in the initial slopes of the loops shown in Fig. 1, as
well as the appearance of a peak in $M$ for applied field values
$H_{\rm a}$ much closer to the central position. We show in Fig. 3b
the calculated field profiles for the same case, $p=10$, as in Fig.
3a, but for a thin disk with $L/R=0.1$. The large effect of
demagnetizing fields is made manifest by the almost constant field
profiles characteristic of high applied fields already achieved for
low values of $\vert H_{\rm a}\vert$. 

This has important consequences for the method used to extract
$J_c(\vert H_{\rm i}\vert)$ from the width of the magnetization loop
$M(H_{\rm a})$. As discussed by Chen and Goldfarb \cite{chengold}
there are two requirements for using the method: (1) The
magnetization on ascending and descending branches of the hysteresis
loop at a given field $H_{\rm a}$ must correspond to fully penetrated
states; and (2) the maximum deviation of $J_c(\vert H_{\rm i}\vert)$
in the sample from the value of $J_c(\vert H_{\rm i}\vert)$ for
$H_{\rm i}=H_a$ must be small. Condition (1) is easily fulfilled as
long as one takes care to measure a reverse magnetization curve
starting from a sufficiently large maximum applied field value. The
key condition is then the second, which is understandable taking into
account the fact that the formula for $J_c$ extraction is based on
the Bean's model for constant $J_c$ so large inhomogenieties in the
local currents will yield wrong results. Comparing our results for
field profiles calculated for the same material but with different
dimensions, see Figs. 3a and 3b, one can clearly see that the field
profiles in the case of thin samples are much more spatially uniform
than in the case of long superconductors. (Actually, the value of the
internal field for thin samples is basically equal to the external
$H_{\rm a}$ value except within a narrow region around $\vert H_{\rm
a}\vert=0$.) There can be several reasons for that. Currents are low,
except when $\vert H_{\rm a}\vert$ is very small, owing to the
contribution of the demagnetizing fields to $H_{\rm i}$ in the
$J_c(\vert H_{\rm i}\vert)$ function, producing small contribution to
the total field. Moreover, in a thin disk a much smaller
total current is flowing as compared with a bulk cylinder. Therefore,
thin samples in transverse geometry are an optimum case for obtaining
$J_c$ from the width of the magnetization loop. This conclusion is
confirmed in Fig. 4, where we compare the $J_c(H_{\rm a})$ function
extracted from the width of the loop with the intrinsic $J_c(\vert
H_{\rm i}\vert)$, for the cases $p=3$ and 10, and for a long sample
($L/R=10$) and a thin one ($L/R=0.1$) for each case. The agreement
between both $J_c$ functions is clearly better for the thin samples
than for the long ones for all
magnetic field values, and particularly for the low-field region
which involves in the long sample a large measurement error. Our
model also shows that for higher values of $p$ the agreement between
$J_c$ functions is better, which implies that measuring thin
superconductors is specially important when the expected dependence
of $J_c$ on $\vert H_{\rm i}\vert$ is strong.

We conclude from this analysis of demagnetizing effects in realistic
finite superconductors that the extraction of the critical current
density from magnetization measurements is best done in thin film
geometry in perpendicular field, or, if thin films are not available,
with the field applied along the sample shortest dimension. In this
way, the superconductor critical current
density and its dependence on the internal field can be precisely
obtained by measuring the hysteresis loops of superconductors using
the non-destructive, reproducible, and widespread magnetization
measurements.

We thank Ministerio de Ciencia
y Tecnolog\'\i a project number BFM2000-0001, and CIRIT project
number 1999SGR00340 for financial support.

\begin{figure}
\caption{
Magnetization loops for the exponential
dependence of the critical current, for (a) $p=0$ (Bean model),
(b) $p=3$, and (c) $p=10$, and for different values of the
length-to-radius, $L/R$ = $\infty$ (solid), 1 (dashed), and 0.1
(dotted).
}
\end{figure}

\begin{figure}
\caption{Calculated applied field at which a peak in the reversal
magnetization of superconducting cylinders occurs, $H_{\rm peak}$, as
a function of $L/R$. Filled circles correspond to $p=3$ and open
circles to $p=10$. Horizontal lines represent the known values for
infinite cylinders.
}
\end{figure}

\begin{figure}
\caption{Internal magnetic field $H_{\rm i}$ profiles in the midplane
of the superconductor cylinder for the case
$p=10$ and $L/R =$ (a) 10 and (b) 0.1, corresponding to the reverse
curve after a maximum applied field of $H_{max}=1.5 H_p$. The
values of the applied field $H_{\rm a}$ range from 1.4$H_p$ to
-1.4$H_p$ in steps of 0.2$H_p$ (from top to bottom). 
}
\end{figure}

\begin{figure}
\caption{Critical current density calculated
from the width of the magnetization loop $\Delta M$ using Eq. (2) for
$L/R$=0.1
(dashed line), and for $L/R=10$ (dotted line), and from the
analytical expression of the exponential dependence (solid line). The
case (a) corresponds to
$p=10$ and (b) to $p=3$. $H_a$ is
the external applied field at which $\Delta M$ is
evaluated, whereas $H_{\rm i}$ is the internal field upon which $J_c$
depends.}
\end{figure}

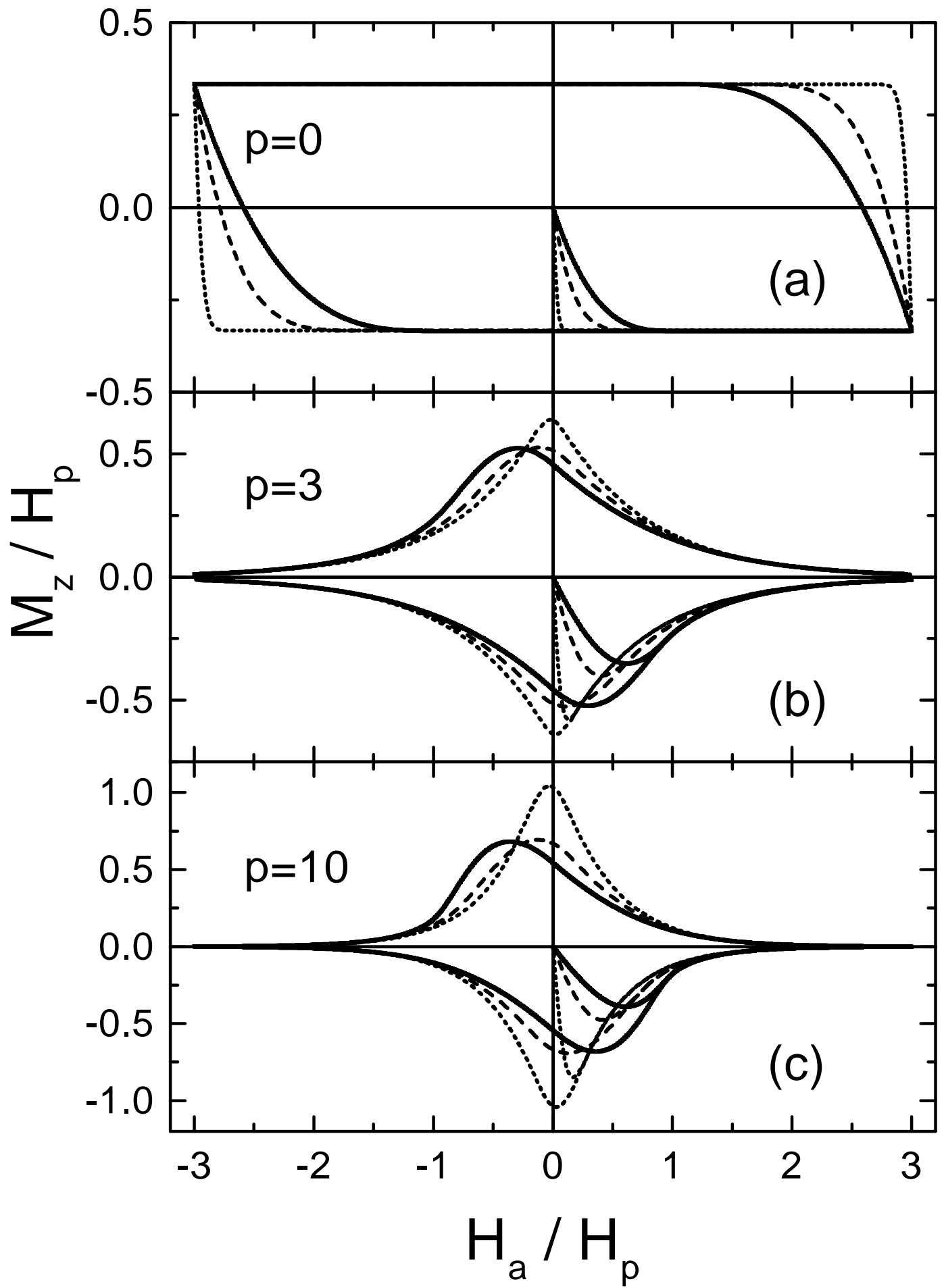
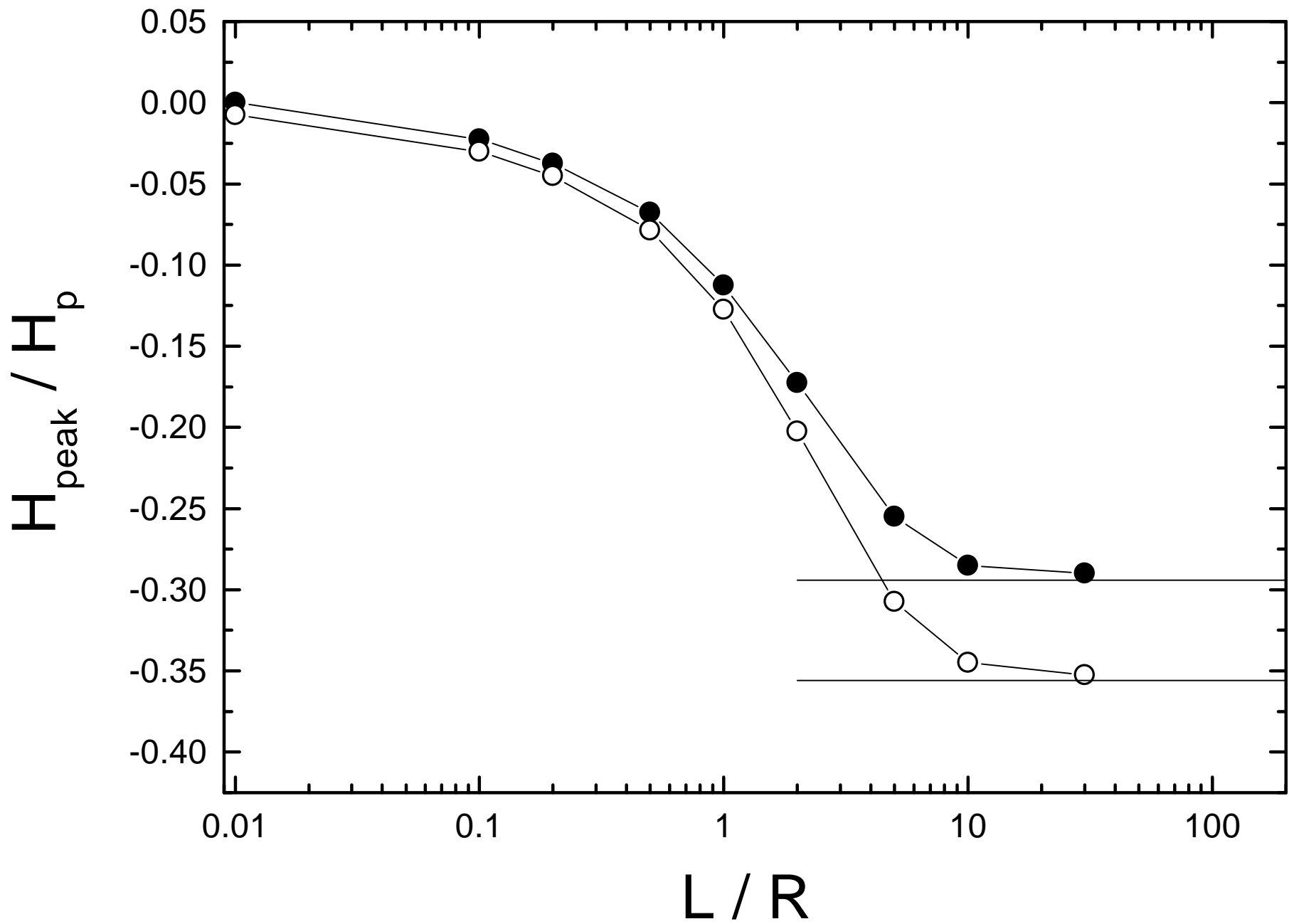
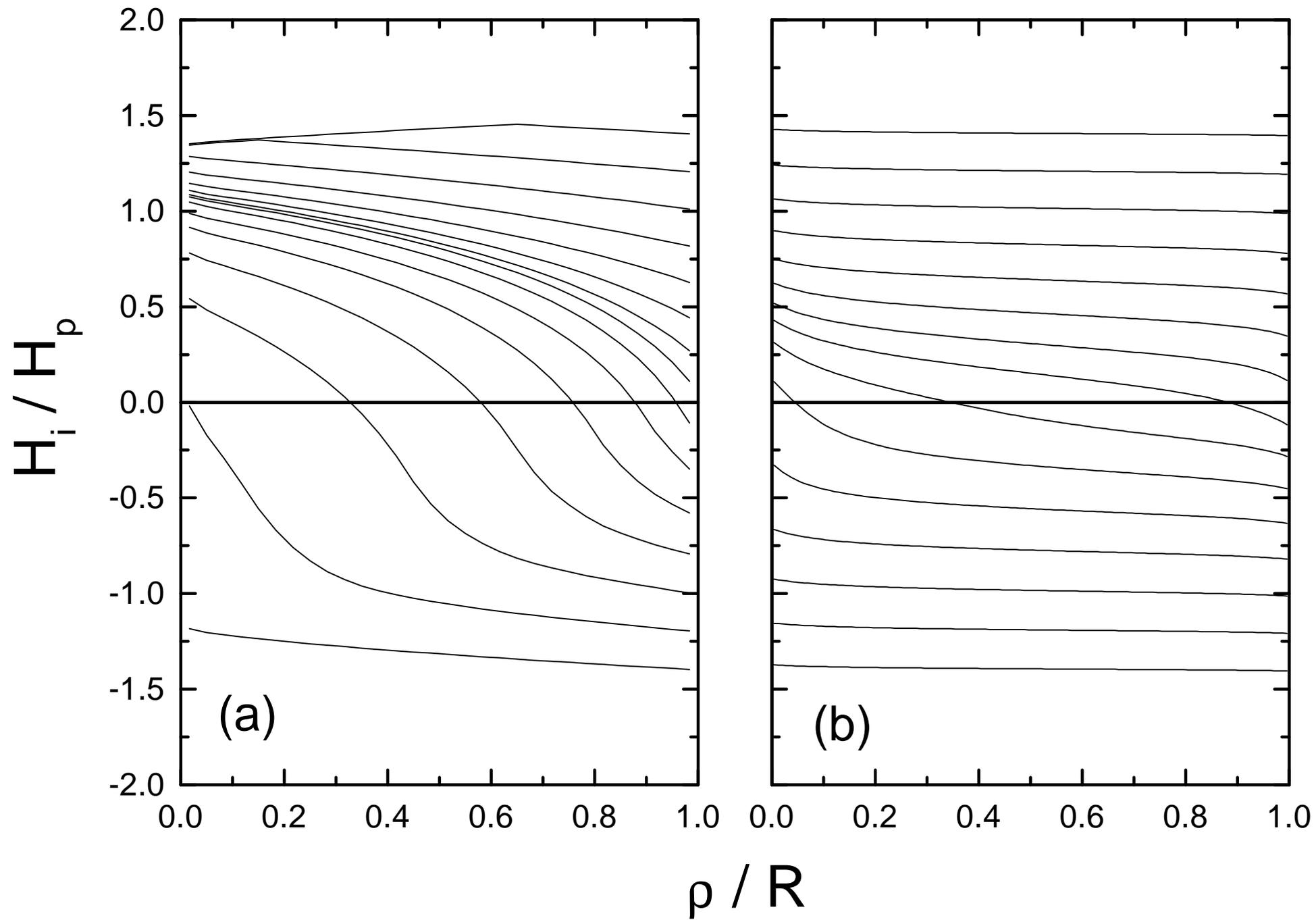
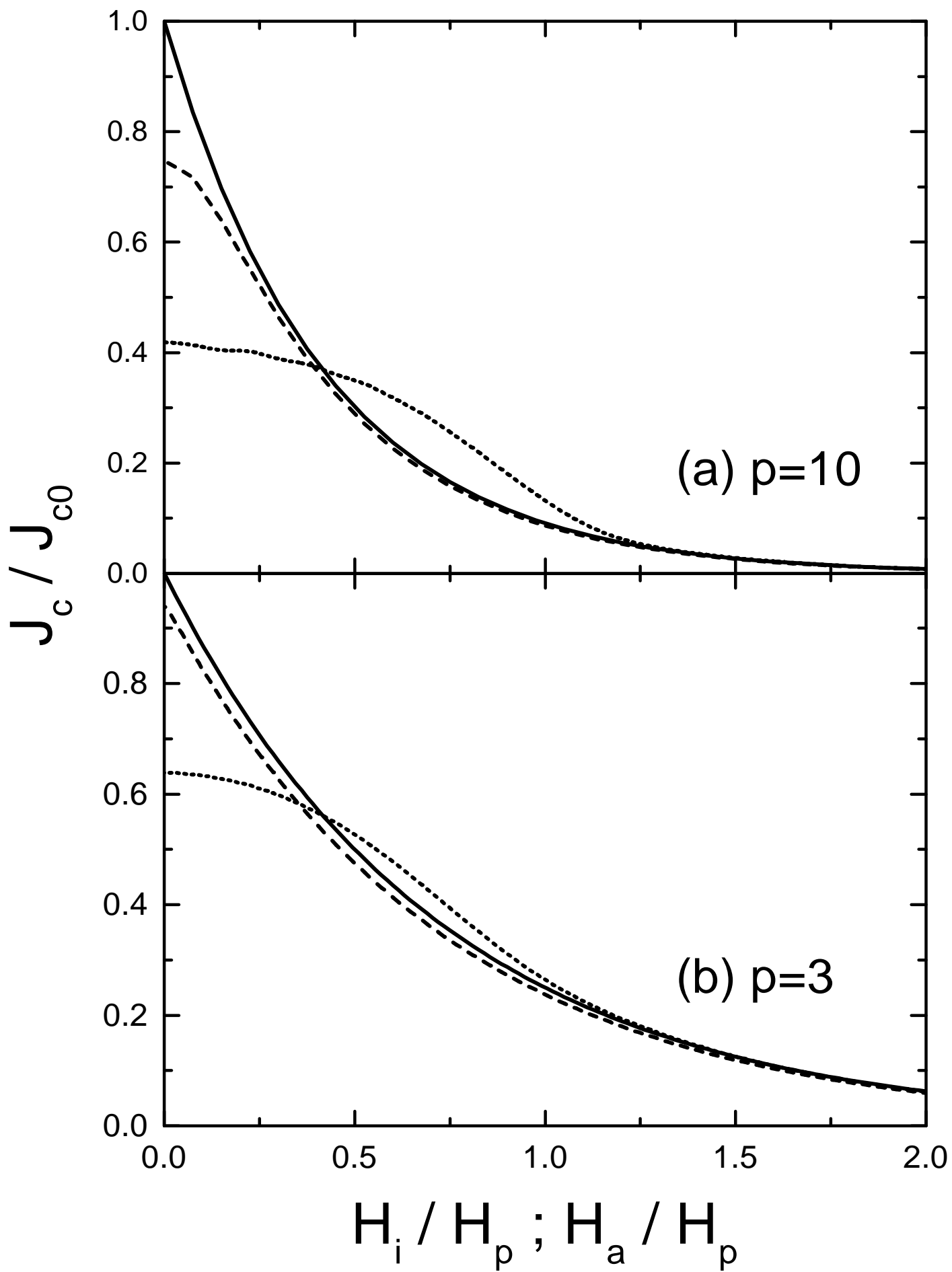
\end{document}